# Synthetic biology: From a word to a world[1]

Xiaojun Hu[*] and Ronald Rousseau[**]

[*]*xjhu@zju.edu.cn*
Medical Information Centre, Zhejiang University School of Medicine, Hangzhou, 310058 (China)

[**]*ronald.rousseau@uantwerpen.be*
Institute for Education and Information Sciences, IOIW-IBW, University of Antwerp (UA), Antwerp, B-2000 (Belgium) & KU Leuven, Dept. Mathematics, Leuven, B-3000 (Belgium)

## Introduction

Synthetic biology can be defined as the application of engineering principles to the fundamental components of biology. More precisely the UK Royal Society (Royal Society, s.a.) describes synthetic biology as follows:

> *Synthetic biology is an emerging area of research that can broadly be described as the design and construction of novel artificial biological pathways, organisms or devices, or the redesign of existing natural biological systems.*

The term synthetic biology was first introduced by the French scientist Stéphane Leduc (1912), be it with a different meaning as today's, and (it seems) in modern times by the Polish geneticist Waclaw Szybalski (1974). Putting aside priority questions it is true that the term gained popularity in mainstream science only in the year 2004 when the first international meeting, called Synthetic Biology 1.0, was held at MIT. Envisaging loads of other applications scientists nowadays declare that they can do better than evolution (Schuster, 2013). Among other aspects, Schuster points to promising aspects for information storage, recalling a pilot study (Church et al., 2012) in which an entire book was stored on a single DNA molecule.

The purpose of this investigation is to reveal the what – where - when of the current situation in this emerging field. In particular, we calculated "year based h-type indices" for high-frequent keywords.

**Literature review**

The main article using informetric techniques to study the field of synthetic biology is (Oldham, Hall and Burton, 2012). They explore the field to inform debates on the governance (related to the United Nations Convention on Biological Diversity) of the field. For this reason they focus on different visualizations of the field. Based on WoS data they distinguish between two groups of articles: the core consisting of 1.255 publications and a group of articles citing the core leading to another 5.995 items. Their core was obtained by a topic search for "synthetic biology", "synthetic genomics", "synthetic genome" or "synthetic genomes". Details are discussed further on when comparing their results with ours. We note though that Oldham et al. (2012) observed the incipient diversification of synthetic biology into several subspecialties. They point out that taking this diversification into account is important for policy debates as synthetic biology may cease to be a 'unitary' object for policy action.

---

[1] This work was partly supported by the National Natural Science Foundation of China (NSFC Grant No 71173185). The authors would like to thank Zhu Zhaoyang of Zhejiang University School of Medicine for helping with data extracting.

Recently Goldman (2014) studied the related field of systems biology, using this field as an example to explore changes in the disciplinary structure of a field. She works under the assumption that concepts from systems biology are transmitted by papers linked via journals to various disciplines. Using a bipartite network she explores connectivity among subject categories and journals. Over the period 2000-2011 the number of subject categories and the number of journals both increased, while the percentage of subject categories with betweenness centrality equal to zero decreased. The whole structure can be characterized as a core - semi-periphery - periphery structure. She further notes that growth at the periphery occurs largely through interdisciplinary journals. Her time study reveals that several clinical disciplines such as Immunology and Oncology move toward the core over time.

## *Methods*
*Data collection*
We employed the following methodology to obtain the data for our investigation.
(1) Essential records were retrieved using the term "synthetic biology" as a topic search in the Web of Science (WoS):
TS="synthetic biology" and document type = article
Databases=SCI-EXPANDED, SSCI, A&HCI, CPCI-S, CPCI-SSH Timespan=2000-2013
This search retrieved 1333 records (retrieved date: Jan.11, 2014).
(2) Next, we extracted the "Keywords" and "Keywords Plus" from all 1333 records, and obtained their frequencies. In this way 6054 terms were found and ranked by frequency.
(3) We sought the precise meanings of the terms in this list making use of MeSH (Medical Subject Headings) definitions. This led to a list of most used content terms related specifically to synthetic biology (recall that one of the authors is head of the Medical Information Center of Zhejiang University School of Medicine, and hence is capable of making this kind of evaluation).
(4) We then used these specific content terms related to synthetic biology to expand the original query, leading to the final search string:
TS= ("synthetic biology " or "synthetic gene network*" or biobrick* or "protein design*" or "genetic circuit*" or "gene regulatory network*" or "cell-free protein synthes*" or "metabolic engineering" or "protein engineering" or "promoter engineering" or "DNA assembly" or "RNA engineering biosensors" or "multipart DNA assembly" or "sequential circuits" or "benchmark synthetic circuits" or "DNA nanotechnology" or "human artificial chromosome" or "synthetic promoters" or "transcriptional circuits" or "abstract genetic regulatory network*" or "gene assembly" or "post-transcriptional regulation" or "engineered proteins" or "cell-free gene circuits") AND Document Types=(Article)
Databases=SCI-EXPANDED, SSCI, A&HCI, CPCI-S, CPCI-SSH Timespan=2000-2013

In this way, 13,836 records were obtained (retrieved date: Jan.15, 2014). This is the set discussed in this article. We note that Goldman (2014) only used the topic search term "systems biology" and included all publication types, leading to 4,446 publications over the period 2000 through 2011.

As "synthetic biology" is said to hold great promise for commercialization we also performed a search for patents in the Derwent Innovations Index (DII), using a similar search query as in the WoS. The search was performed on January 24, 2014 and this for the Timespan=2000-2013. This resulted in 788 patent records.

*Data processing*
(1) Topic keyword counting. We determined the keyword frequency based on the retrieved 13,836 records, and their yearly distribution.
(2) Dynamical changes in keyword use. To find out dynamical changes over the period 2000-2013, we calculated "year based h-type indices" for the high-frequent keywords (Mahbuba & Rousseau, 2013).

## *Basic results*

In this section we show basic results: most active countries/regions and organizations; WoS categories and areas to which articles on synthetic biology belong; number of articles per year and growth of the field (using WoS' *analyze* functionality). We abbreviate the term Synthetic Biology referring to the set of articles retrieved by our query as SB.

The most-active countries/regions over the period [2000 – 2013] are shown in Table 1. Recall that the WoS assigns an article to each country with at least one participating author as shown by the institutional address. Besides rankings over the whole period we also show number of publications and rankings for the first and the second half of the period. Moreover, we calculated the percentage of articles about synthetic biology among all articles (by that country, over the same period) and the ranking (restricted to the 20 countries/regions studied here) according to this parameter.

Table 1. Most-active countries/regions over the period [2000 – 2013], data from the WoS.

| Rank | Country | # articles | # publications and ranking [2000 – 2007] | # publications and ranking [2008-2013] | % SB among all articles and ranking |
|---|---|---|---|---|---|
| 1 | USA | 5973 | 2419 (1) | 3554 (1) | 0.144 (2) |
| 2 | GERMANY | 1392 | 504 (3) | 888 (3) | 0.129 (5) |
| 3 | JAPAN | 1294 | 630 (2) | 664 (4) | 0.125 (7) |
| 4 | PEOPLES R CHINA | 1258 | 263 (5) | 995 (2) | 0.092 (13) |
| 5 | ENGLAND | 966 | 347 (4) | 619 (5) | 0.101 (12) |
| 6 | FRANCE | 621 | 244 (6) | 377 (6) | 0.080 (16) |
| 7 | CANADA | 553 | 224 (7) | 329 (8) | 0.088 (14) |
| 8 | SOUTH KOREA | 508 | 164 (9) | 344 (7) | 0.119 (8) |
| 9 | ITALY | 442 | 182 (8) | 442 (10) | 0.074 (17) |
| 10 | SPAIN | 410 | 128 (12) | 282 (9) | 0.083 (15) |
| 11 | NETHERLANDS | 385 | 164 (9) | 221 (11) | 0.110 (10) |
| 12 | SWITZERLAND | 346 | 143 (11) | 203 (12) | 0.136 (3) |
| 13 | AUSTRALIA | 301 | 109 (14) | 192 (14) | 0.069 (19) |

| 14 | INDIA | 300 | 105 (15) | 195 (13) | 0.069 (20) |
| 15 | SWEDEN | 279 | 110 (13) | 169 (15) | 0.113 (9) |
| 16 | DENMARK | 208 | 83 (16) | 125 (18) | 0.149 (1) |
| 17 | ISRAEL | 195 | 57 (18) | 138 (17) | 0.129 (4) |
| 18 | TAIWAN | 184 | 41 (19) | 143 (16) | 0.070 (18) |
| 19 | SCOTLAND | 158 | 35 (20) | 123 (19) | 0.106 (11) |
| 20 | FINLAND | 156 | 75 (17) | 81 (20) | 0.126 (6) |

China, South Korea and Taiwan moved up in the rankings when comparing the second period with the first one. Among the top countries Japan lost in the rankings. The ranking according to the percentage of articles devoted to Synthetic Biology shows that, on the one hand, Denmark, Israel and Finland have a high percentage of articles on SB. On the other hand China, although ranking second in the second period is only 13th in the ranking per percentage devoted to SB, illustrating the fact that China has many other priorities. Also Canada, France, Italy and Spain have other priorities. Compared with the results of Oldham et al. (2012) we notice several differences: UK is second in their core group, Switzerland 5th, Spain 6th, Japan 8th and China 10th. Yet, in the citing articles group China becomes 4th.

Divided over continents Europe and North America have an equal share (37%), followed by Asia (22%), Latin America (2%) and Oceania (2%). Africa's share is below 1%.

Most active organizations are shown in Table 2. This list is clearly dominated by American universities. Yet, this list has no clear top university or small group of top organizations but numbers decrease slowly. We further note that the first company in this list is Genentech Inc. on rank 185 with 27 articles. This seems to indicate that, although synthetic biology can be considered an applied field it is not yet a field which is ripe for large scale commercialization.

Table 2. Most-active organizations.

| Organization | # articles |
|---|---|
| MIT (USA) | 244 |
| CHINESE ACAD SCI (P.R. China) | 242 |
| HARVARD UNIV (USA) | 242 |
| CALTECH (USA) | 228 |
| STANFORD UNIV (USA) | 221 |
| UNIV TOKYO (JPN) | 203 |
| UNIV CALIF BERKELEY (USA) | 197 |
| DUKE UNIV (USA) | 148 |
| UNIV WASHINGTON (USA) | 147 |
| UNIV ILLINOIS (USA) | 138 |

Again Oldham et al. (2012) obtain different results. Their list of most-active organizations consists of the University of California Berkeley, the Swiss Federal Institute of Technology (ETH), Harvard and MIT. We found 121 articles for ETH. Clearly, as already shown on country level, China and Japan are underrepresented in their investigation.

Delving somewhat deeper into this we also performed a search for patents in the Derwent Innovations Index (DII), using a similar search query as in the WoS. Contrary to article publishing institutions, patent assignees are mostly Japanese and Korean. Yet numbers of assigned patents are an order of magnitude less than numbers of publications, affirming the observation that the field is not yet ripe for large-scale commercialization.

Table 3. Most-active assignees (from the DII search).

| Assignee Name | # patents |
|---|---|
| CELLFREE SCI KK (=CO LTD) (JP) | 25 |
| MACROGEN CO LTD (Korea) | 23 |
| DOKURITSU GYOSEI HOJIN RIKAGAKU KENKYUSH (JP) | 17 |
| SHIMADZU CORP (JP) | 17 |
| TOYOBOSEKI KK (JP) | 17 |
| MASSACHUSETTS INST TECHNOLOGY (MIT) (USA) | 12 |
| NEC ELECTRONICS CORP (JP) | 12 |
| UNIV LOUISIANA STATE & AGRIC & MECH COLL (USA) | 10 |
| RIKEN KK (JP) | 10 |
| UNIV CALIFORNIA (USA) | 9 |

The multidisciplinary aspects of SB are clearly shown by the WoS categories involved in its research. Table 4 shows the top-5 categories. Also Oldham et al. (2012) have Biochemistry & Molecular Biology as leading subject category (core and citing articles), followed by Chemistry (for the citing articles group) and Biotechnology & Applied Microbiology (second in the core). Differences between our results and Oldham et al.'s or Goldman's are due to the methodology. In particular, we used a more inclusive search query than our colleagues.

Table 4. WoS categories most involved in SB research.

| WoS categories | % of all articles |
|---|---|
| BIOCHEMISTRY & MOLECULAR BIOLOGY | 31.9 |
| BIOTECHNOLOGY & APPLIED MICROBIOLOGY | 21.7 |
| MULTIDISCIPLINARY SCIENCES | 9.1 |
| BIOCHEMICAL RESEARCH METHODS | 8.0 |
| MATHEMATICAL & COMPUTATIONAL BIOLOGY | 6.7 |

Research in SB is often supported by grants from large funding bodies. The WoS yields a list of 8,455 names, be it that there are many funds occurring under several names. Table 5 shows the most-important ones: NIH USA has more than 1000 supported articles, while the other ones have each at least 200 supported articles.

Table 5. Most important funding organizations.

| |
|---|
| National Institutes of Health (NIH) - USA |
| National Science Foundation USA |
| National (Natural) Science Foundation China |
| European Union (EU) / European Commission (EC) |
| Deutsche Forschungsgemeinschaft (DFG) |

Oldham et al.'s list of funding institutes is dominated by the NIH, NSF (USA) and the European Programs. Again, China's research (funded by NSFC) is underrepresented.

Doing better than evolution has a touch of "playing god" and certainly entails moral obligations and ethical problems. Adding the topic terms "ethic*"OR "moral*" to the main query led to 54 articles. The largest group (17) belongs to the WoS Category Ethics, followed by Social Sciences Biomedical (12). More than half were published in the latest two years.

*Growth in the number of articles on synthetic biology*

The yearly growth curve is shown in Figure 1. This curve can best be described as exponential growth. Giving the year 2000 the x-value 0 (and hence 2013 the x-value 13) a best-fitting curve is given by $y = 454.3 \, e^{0.105x}$ ($R^2 = 0.97$), where y denotes the yearly number of published articles on SB.

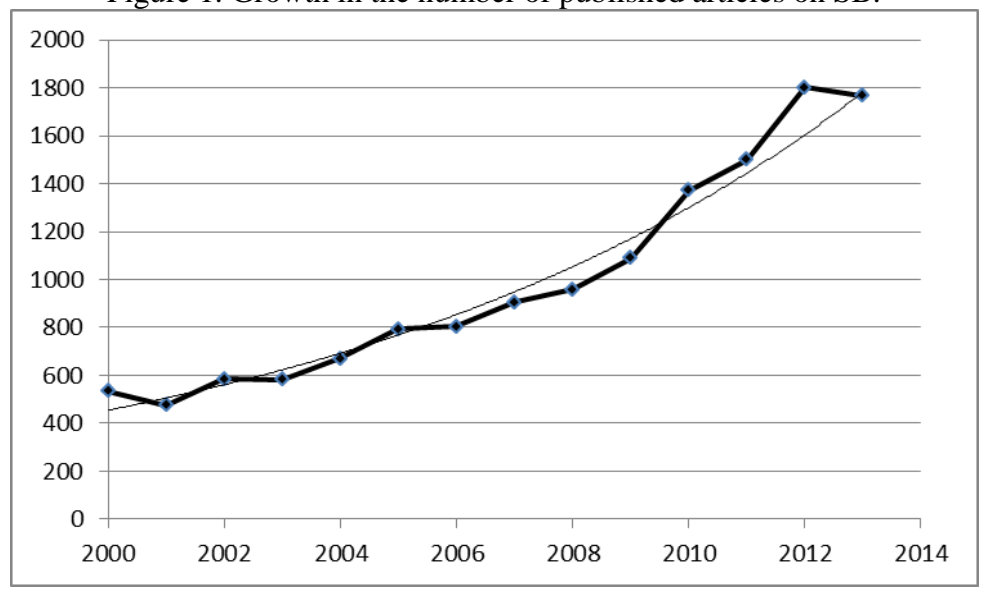

Figure 1. Growth in the number of published articles on SB.

## *Topic keywords and year-based h-indices*

We found a total of 22,253 keywords (not including Keywords Plus, as these were generally too broad) in the retrieved records. However, the majority of them (76%) occurred just once, reflecting the broadness of the field, as well as the fact that, being in an emerging stage, terminology has not yet settled. Remarkably, the term synthetic biology (and related terms) occurred just 28 times (period 2000-2013) proving that we had to look into the field's "world" rather than just considering the "word". Focusing on major topics we brought keywords and related forms together under one name. In this way we obtained 35 high-frequent topic keywords each occurring at least 100 times. We removed general topics such as cell, enzyme, genetic, gene, protein, E. coli and their related terms, leading to 28 keywords, representing the hot topics in the field of SB. These keywords were analysed using a recently introduced approach based on year-based h-indices (Mahbuba & Rousseau, 2013).

We recall the following definitions. Consider a given topic term T and assume that years (here restricted to the period 2000-2013) are ranked according to the number of articles published dealing with this topic. Then this topic's year-based h-index is equal to t if t is the highest rank such that in the first t years t articles were published dealing with this topic. Let Z and Y be the latest and the oldest years included in the topic's h-core, then the period [Y, Z] is called the core interval. If Z-Y+1 = t then there is no gap in the core. The core gap is defined as Z-Y+1-t, or informally: the number of missing years in the core. Finally, the relative core gap for topic t is defined as: (core gap / t). In (Hu & Rousseau, 2014) we have shown how using these notions may provide an easy-to-use overview of a field. Table 6 shows the results of this analysis for the SB set.

Table 6. Hot topic keywords in the research field of synthetic biology and their year-based activity h-type indices (period 2000-2013).

| Topic keywords | Year-based h-index | Core interval | Top year | Core gap | Relative core gap |
|---|---|---|---|---|---|
| protein engineering | 14 | 2000-2013 | 2011 | 0 | 0 |
| metabolic engineering | 14 | 2000-2013 | 2013 | 0 | 0 |
| protein design+* | 14 | 2000-2013 | 2012 | 0 | 0 |
| DNA+ | 11 | 2002-2013 | 2013 | 1 | 0.09 |
| microRNA+ | 10 | 2002-2013 | 2013 | 2 | 0.2 |
| cell-free protein synthesis+ | 10 | 2004-2013 | 2007 | 0 | 0 |
| protein folding+ | 10 | 2000-2011 | 2004 | 2 | 0.2 |
| RNA+ | 9 | 2003-2013 | 2012 | 2 | 0.22 |
| mutagenesis+ | 9 | 2004-2013 | 2002 | 1 | 0.11 |
| gene expression+ | 9 | 2002-2013 | 2013 | 3 | 0.33 |
| stability+ | 9 | 2004-2013 | 2013 | 1 | 0.11 |
| fluorescence+ | 9 | 2001-2013 | 2013 | 4 | 0.44 |
| protein stability+ | 9 | 2001-2013 | 2004 | 4 | 0.44 |
| gene regulatory network+ | 8 | 2006-2013 | 2012 | 0 | 0 |
| directed evolution+ | 8 | 2005-2013 | 2012 | 1 | 0.125 |
| nano+ | 8 | 2005-2013 | 2013 | 1 | 0.125 |
| evolution+ | 8 | 2003-2013 | 2013 | 1 | 0.125 |
| systems biology+ | 8 | 2004-2013 | 2012 | 2 | 0.25 |
| microarray+ | 8 | 2006-2013 | 2012 | 0 | 0 |
| sequential circuit+ | 8 | 2000-2008 | 2000 | 1 | 0.125 |
| biocatalysis+ | 7 | 2005-2013 | 2013 | 2 | 0.29 |
| combination+ | 7 | 2002-2013 | 2013 | 5 | 0.71 |
| gene regulation+ | 7 | 2005-2013 | 2013 | 2 | 0.29 |
| self-assembly+ | 7 | 2007-2013 | 2012 | 0 | 0 |
| antibody+ | 7 | 2002-2013 | 2013 | 5 | 0.71 |
| dynamics+ | 7 | 2007-2013 | 2012 | 0 | 0 |
| protein-protein interaction+ | 7 | 2006-2013 | 2013 | 1 | 0.14 |
| genome+ | 6 | 2006-2013 | 2012 | 2 | 0.33 |

*The symbol "+" indicates a keyword and its related terms

Clearly, protein engineering, metabolic engineering and protein design are the overall hot topics in synthetic biology. Oldham et al. (2012) found the following top terms: synthetic biology, E. coli (a term we removed), gene expression, systems biology and metabolic engineering. Table 6 shows that the core interval of most topics extends to the latest year (2013). Moreover the top year (year in which the most articles on this topic were published) is often 2012 or 2013, indicating that the interest in these topics is still growing. Interest in sequential circuits seems to have passed its peak. We refer a more detailed discussion of the dynamics in the use of these topics to the conference presentation.

## *Conclusion*

Clearly synthetic biology is one of the battlefields where the main countries fight for the supremacy in science (Joyce et al., 2013). The word "synthetic biology" hides a "big world", ready to be explored by interdisciplinary research collaborations. We hope that our informetric study brings a new perspective to the study of this innovative field.